\documentclass[pdftex,preprint,epjc3]{svjour3}          
\pdfoutput=1
\RequirePackage[T1]{fontenc}

\smartqed  
\RequirePackage{graphicx}
\RequirePackage{mathptmx}      
\RequirePackage{flushend}
\RequirePackage[numbers,sort&compress]{natbib}
\RequirePackage[colorlinks,citecolor=blue,urlcolor=blue,linkcolor=blue]{hyperref}

\newcommand{\prl}[3]{ Phys.\ Rev.\ Lett. {\bf #1}, #3 (#2)}
\newcommand{\prd}[3]{ Phys.\ Rev.\ D {\bf #1}, #3 (#2)}
\newcommand{\prc}[3]{ Phys.\ Rev.\ C {\bf #1}, #3 (#2)}
\newcommand{\npa}[3]{  Nucl.\ Phys.\ A {\bf #1}, #3 (#2)}

\newcommand{\plb}[3]{  Phys.\ Lett.\ B {\bf #1}, #3 (#2)}
\newcommand{\epjc}[3]{  Eur.\ Phys.\ J.\ C {\bf #1}, #3 (#2)}
\newcommand{\jhep}[3]{ JHEP {\bf #1}, #3 (#2)}

\journalname{Eur. Phys. J. A}

\begin{document}
\title{A unified approach to hadron phenomenology at zero and finite temperatures in a hard-wall AdS/QCD model}


\author{Zhiyuan Wang\thanksref{e1,addr1}
        \and
        Bo-Qiang Ma\thanksref{e2,addr1,addr2,addr3} 
}

\thankstext{e1}{e-mail: lagrenge@pku.edu.cn}
\thankstext{e2}{e-mail: mabq@pku.edu.cn}

\institute{School of Physics and State Key Laboratory of Nuclear Physics and Technology, Beijing 100871, China\label{addr1}
          \and
          Collaborative Innovation Center of Quantum Matter, Beijing, China\label{addr2}
          \and
          Center for High Energy Physics, Peking University, Beijing 100871, China\label{addr3}
}

\date{Received: date / Accepted: date}

\maketitle

\begin{abstract}
We propose a unified approach to study meson, nucleon and $\Delta$-baryon properties at zero and finite temperatures in the context of hard-wall AdS/QCD model. We first combine some previous works dealing with mesons and baryons separately, and introduce a new parameter~$\xi$ so that the model could give a universal description of spectrum and couplings of both sectors in a self-consistent way. All observables calculated numerically show reasonable agreement with experimental data. We then study these observables at nonzero temperature by modifying the AdS space-time into AdS-Schwartzchild space-time. Numerically solving the model, we find an interesting temperature dependence of the spectrum and the couplings. We also make a prediction on the finite temperature decay width of some nucleon and $\Delta$ excited states.
\end{abstract}

\section{Introduction}
\label{sec:intro}
Quantum chromodynamics~(QCD) is now generally believed to be the basic theory of strong interactions. Due to asymptotic freedom, it can describe the strongly interacting particles at high energy where perturbative expansions are valid. In the low energy region, however, QCD becomes strongly coupled and highly nonlinear, and has thus far precluded any analytic solutions. How to study the low energy dynamics of QCD is still a difficult problem in physics.

The Anti-de Sitter/conformal field theory~(AdS/CFT) correspondence, first proposed by Maldacena~\cite{Maldacena} in 1998, is originally intended to study the strong coupling region of a particular conformal gauge theory--the ${\cal N}=4$ supersymmetric Yang-Mills~(SYM) theory, by means of its duality to a weakly coupled string theory in AdS$_5\times$S$_5$, and has later on been extended to many other gauge theories~(for a review, see Ref.~\cite{AdS_CFT_review}). Although at first sight, QCD differs a lot from ${\cal N}=4$ SYM, arguments have been given~\cite{QCD_conformal,Brodsky:2014yha} that in the small momentum transfer region, where the effective QCD coupling is approximately constant, QCD resembles a conformal gauge theory, and can thus be studied using a gauge/gravity type model. In this approach, one introduces an infrared~(IR) cut-off in the fifth dimension of the AdS space to model QCD confinement, and then introduces some five-dimensional~(5D) fields corresponding to boundary four-dimensional~(4D) operators. The correlation functions of boundary operators can be obtained by functionally differentiating the on-shell 5D action (following the AdS/CFT prescription), and all the physical quantities are calculated from the correlation functions through the Lehmann-Symanzik-Zimmermann~(LSZ) formalism. This approach is currently referred to as ``bottom-up'' AdS/QCD correspondence. Although it has not been strictly proved yet, it has turned out to be phenomenologically successful in obtaining hadronic bound state properties, such as meson spectrum~\cite{meson_spectra_Brodsky,meson_spectra_Erlich,Gutsche:2013prd} and form factors~\cite{QCD_conformal,meson_EMFF,meson_TFF,PiPiFF}.

The AdS/CFT correspondence can also be modified to study finite-temperature field theories~\cite{witten}. In this approach, one introduces a Schwartzchild blackhole in the AdS$_5$ background and it turns out that the resulting 5D gravity theory corresponds to a boundary field theory at temperature $T=1/(8\pi M)$, where $M$ is the mass of the blackhole. This idea was later applied  to AdS/QCD to study the temperature effects on meson spectrum~\cite{meson_T,vmeson_T}, heavy quark potentials~\cite{Heavy1,Heavy2}, and the confinement-plasma phase transition~\cite{QGP_trans1,QGP_trans2}.

While the holographic model of the meson sector has received phenomenological success, the extension to the nucleon sector raises a problem. Imitating the meson sector, we can similarly introduce the 5D Lagrangian and then solve the eigenvalue problem to obtain nucleon spectrum~\cite{baryon_spectra,Zhangpeng}, but it turns out that the value of the IR cut-off $z_m$ needs to be 1.5--3 times the value used in meson sector. We can just use a different $z_m$ for the nucleon sector and set mesons aside since there is no reason why mesons and nucleons must have the same confinement scale, but the consequence is that the definition of the covariant derivative on fermion fields becomes problematic, making it difficult to study the meson-nucleon coupling constant $g_{\pi NN}$. There is an attempt to deal with this problem using the same parameter $z_m$~\cite{unifyapproach}, with numerical error 10\%--20\% in the spectrum and $50\%$ in coupling constant $g_{\pi NN}$. In this paper we make two different attempts to this issue. The first one is to use different $z_m$ for the two sectors but redefine the interaction Lagrangian in a self-consistent way so that the coupling $g_{\pi NN}$ is uniquely determined. The numerical result of both the spectrum and coupling constants are in reasonable agreement with experiment. The second one is to use the same $z_m$ but slightly modify the AdS/QCD prescription, introducing a ``magnification factor'' $\xi$ for the nucleon sector to indicate a different confinement scale. The numerical results prove to be the same, so we put the latter one in the appendix.

The problem addressed in the previous paragraph also presents in the spin-3/2~($\Delta$-baryon) sector. A previous work~\cite{S23} has established the necessary formalism of studying spin-$3/2$ baryons in AdS/QCD, with the error of the calculated spectrum being 13\% to 27\%. In this paper we use the same technique mentioned in the previous paragraph with the same $\xi$ parameter to get a better fit with experimental data. The calculated spectrum fits well, and the coupling constant $g_{\pi\Delta\Delta}$ is consistent with experimental data and the prediction of other theoretical approaches.

The previous work involving $\Delta$ baryons focused on the zero temperature case. However, it is increasingly important nowadays to study $\Delta$ baryons at finite temperature. We know that $\Delta$ baryons carry spin $3/2$ and isospin $3/2$, they often arise as intermediate resonance states of nucleon-pion scattering. Thus a knowledge of their finite-temperature properties would be important for studying thermal nucleon-pion reactions that occur in nuclear environments, such as in Relativistic Heavy Ion Collider~(RHIC) experiments~\cite{RHIC_1997,STAR_2004}.

Therefore, another aim of this paper is to study the spectrum and effective couplings of $\Delta$ baryons at finite temperature, by modifying the AdS space-time to AdS-Schwartzchild space-time. To be concise, we use the finite temperature formalism from the very begining, all zero temperature formulae can be obtained by setting $T=0$. Solving the equations numerically, we get the temperature dependence of the mass spectrum and the couplings. The temperature dependence of the mass spectrum shows a novel feature of the $\Delta$ sector compared with those of nucleons and mesons.

Our paper is organized as follows. In Sect.~\ref{sec:description}, we describe our unified AdS/QCD model for mesons, nucleons and $\Delta$ baryons at finite temperature, paying special attention to the baryon sector, and give a universal fitting of all the observables of the three sectors. In Sect.~\ref{sec:finite}, we go to the nonzero temperature region and study the temperature dependence of the spectrum and couplings. Finally, in Sect.~\ref{sec:summary}, we briefly summarize our results.

\section{Description of the model}
\label{sec:description}
We use the following space-time metric for 5D AdS-Schwarzchild~(AdS-Sch) background:
\begin{equation}\label{metric}
\mathrm{d}s^2=\frac{1}{z^2}\left(f^2(z)\mathrm{d}t^2-(\mathrm{d}x^i)^2-\frac{1}{f^2(z)}\mathrm{d}z^2\right), ~~~~\epsilon<z<z_m,
\end{equation}
where $f^2(z)=1-z^4/z_h^4$. The Hawking temperature of the black hole $T_H=1/(\pi z_h)$ is equal to the temperature of the corresponding 4D boundary field theory. The cut-off at $z=\epsilon$ (with $\epsilon\to 0$ implied) corresponds to UV cut-off in QCD, while the hard-wall cut-off at $z=z_m$ corresponds to IR cut-off, $\Lambda_{\mathrm{QCD}}$, to simulate confinement. According to Ref.~\cite{QGP_trans1}, the AdS-Sch background is stable only at temperatures above the deconfinement temperature $T_c=2^{1/4}/(\pi z_m)$, and becomes metastable when $T$ drops below $T_c$. However, as illustrated in Ref.~\cite{v_meson_f}, when the strong interacting plasma formed in heavy ion collisions at a high temperature quickly cools down, it stays in the supercooled~(metastable) phase corresponding to AdS-Sch background. So in this sense it is more practical to study hadron behaviour in AdS-Sch background~(\ref{metric}).\footnote{By the way, as illustrated in Ref.~\cite{spectra_HP_trans}, in the stable phase at temperature below $T_c$~(the thermal AdS background), the hadron properties~(masses, decay constants) do not change when temperature varies.} In this paper we consider the metric~(\ref{metric}) for temperatures $0<T<T_c$. When $T>T_c$, although the AdS-Sch phase becomes stable, our method of solving eigenvalue problems may not work well, and in this case it is more practical to analyse the baryon spectral functions, which is beyond the scope of this paper.

\subsection{The boson sector}
\label{subsec:boson}
Since QCD has a SU(2)$_R\times $SU(2)$_L$ chiral symmetry, we should introduce two 5D gauge fields $L$ and $R$ corresponding to the Noether currents of 4D global chiral symmetry $J_{L,R}^{a,\mu}=\bar{q}_{L,R}\gamma^\mu t^a q_{L,R}$. We also introduce a 5D scalar field $X^{ab}$ corresponding to quark bilinear operator $\bar{q}^a_L q^b_R$ in the boundary QCD.

The gauge-invariant 5D action of this boson sector can be constructed as
\begin{equation}\label{lagrangian}
S_X=\int \mathrm{d}^4x\int_{\epsilon}^{z_m}dz \sqrt{g}\mathrm{Tr}\left[|\nabla_M X|^2+3|X|^2-\frac{1}{4g_5^2}(F_L^2+F_R^2)\right],
\end{equation}
where $F_{L,R}$ is the field strength tensor corresponding to 5D gauge fields $L,R$, $\nabla_M X=\partial_M X-i L_{M}X+iX R_{M}$ is the covariant derivative acting on $X$, and $\sqrt{g}=1/z^5$ is the square root of the metric determinant.

Turning off the fields $L,R$ and solving the equation of motion for $X^{ab}(x,z)=\delta^{ab}X_0(z)$, one gets
\begin{equation}\label{exp:X}
X_0(z)\equiv \frac{v(z)}{2}=\frac{1}{2}m_q z ~ {_2F_1}(\frac{1}{4},\frac{1}{4},\frac{1}{2},\frac{z^4}{z^4_h})+\frac{1}{2}\sigma z^3 {_2 F_1}(\frac{3}{4},\frac{3}{4},\frac{3}{2},\frac{z^4}{z^4_h}),
\end{equation}
where ${_2F_1}$ is the confluent hypergeometric function. At zero temperature $T=0$, the above solution reduces to $X_0(z)=(m_q z+\sigma z^3)/2$. According to the AdS/CFT dictionary, the coefficient of the non-normalizable term $m_q$ corresponds to the boundary value of $X$ which couples to the quark bilinear operator $\bar{q}_L q_R$. We recognize it as the quark mass term which explicitly breaks chiral symmetry. The coefficient of the normalizable term $\sigma$ corresponds to the expectation value of $\bar{q}_L q_R$: $\sigma=\left<\bar{q} q\right>$, which breaks SU(2)$_R\times$ SU(2)$_L$ spontaneously to SU(2)$_V$.

The 5D coupling $g_5$ can be fixed following Ref.~\cite{meson_spectra_Erlich}:
\begin{equation}
g_5=\sqrt{\frac{12\pi^2}{N_c}}=2\pi.
\end{equation}
There are 3 free parameters $z_m$, $m_q$ and $\sigma$ in this sector,  we use $z_m=1/(346~\mathrm{MeV}), \sigma=(308~\mathrm{MeV})^3,m_q=2.30~\mathrm{MeV}$, which reproduce well 7 experimental observables in this sector~(see Table~\ref{table:summary} in Sect.~\ref{subsec:sumzero}).

The spectrum of the boson sector at zero and nonzero temperature are studied in Ref.~\cite{meson_spectra_Erlich} and Ref.~\cite{meson_T} respectively. The pion 5D wave functions $\pi(z)$ and $\phi(z)$ are defined by
\begin{eqnarray}\label{def_phi_pi}
A_\mu&=&(L_{\mu}-R_{\mu})/2=A_{\mu\perp}+\partial_\mu \phi\nonumber\\
X(x,z)&=&X_0(z)e^{2i\pi^at^a}.
\end{eqnarray}

At nonzero temperature, these functions can be obtained by solving the eigenvalue equation that follows from the equations of motion of $X$ and $A_\mu$:
\begin{eqnarray}\label{pion_T}
z\partial_z(\frac{1}{z}\partial_z\phi)&=&g_5^2\frac{v^2}{z^2 f^2}(\phi-\pi),\nonumber\\
m_\pi^2\partial_z\phi &=&g_5^2\frac{v^2 f^2}{z^2}\partial_z \pi.
\end{eqnarray}
The normalization of the 5D wave functions $\phi,\pi$ are fixed by requiring canonical 4D kinetic terms after Kaluza-Klein reducing the 5D action (\ref{lagrangian})
\begin{equation}\label{norm_pi}
\int_{\epsilon}^{z_m}\left[\frac{\phi^\prime(z)^2}{g_5^2 z}+\frac{v^2}{z^3 f^2}(\pi-\phi)^2 \right]dz=1.
\end{equation}

\subsection{Spin-1/2 sector}
\label{subsec:spin12}

Nucleons in AdS/QCD are described by a pair of 5D Dirac spinors $N_1$ and $N_2$~\cite{baryon_spectra,Nucleon_reso}, corresponding to the 4D baryon operators $O_L$ and $O_R$, respectively. The 5D action for $N_1$ and $N_2$ is :
\begin{eqnarray}\label{action_N}
S_{N}=\int \mathrm{d}^4 x\int^{z_B}_{\epsilon}dz &\sqrt{g}&\left[\frac{i}{2}\bar{N}_1 e^M_A\Gamma^A\nabla_M N_1- \frac{i}{2}(\nabla^\dagger_M\bar{N}_1) e^M_A\Gamma^A N_1-m_5\bar{N}_1 N_1\right.\nonumber\\
&+&\left. (1\rightarrow 2 \,\, ,\,\, L\rightarrow R  \,\, ,\,\, m_5\rightarrow-m_5)\right].
\end{eqnarray}
where $\nabla_M$ is the covariant derivative acting on nucleon fields. The value of the 5D mass $m_5$ is given by the AdS/CFT prescription: $m_5=5/2$.

We mention  in the introduction section that a different IR cut-off should be used in the nucleon sector to make the calculated spectrum agree with experiment, we take it to be $z_B=\xi z_m$. The presence of this $\xi$ parameter indicates that the nucleons have a confinement scale different from mesons. A different~(but equivalent) method to introduce this parameter is given in the appendix.

The definition of the covariant derivative $\nabla_M$ deserves more care. The usual definition is
\begin{equation}\label{cov_Dusual}
\nabla_M=\partial_M+\frac{i}{4}\omega^{AB}_M\Sigma_{AB}-iL^a_M(x,z) t^a,\quad \Sigma_{AB}=[\Gamma_A,\Gamma_B]/2i.
\end{equation}
However, as the meson and baryon fields are defined on different ranges, this definition becomes problematic in that it involves $\int_{\epsilon}^{\xi z_m} dz \bar{N}(x,z)e_A^M \Gamma^A L_M(x,z) N(x,z)$. We should modify it to
\begin{eqnarray}\label{cov_D}
\nabla_\mu&=&\partial_\mu+\frac{i}{4}\omega^{AB}_\mu\Sigma_{AB}-iL^a_{\mu}(x,z/\xi) t^a,\nonumber\\
\nabla_5&=&\partial_5+\frac{i}{4}\omega^{AB}_5\Sigma_{AB}-i\frac{1}{\xi}L^a_5(x,z/\xi) t^a.
\end{eqnarray}
It can be checked easily that such a definition still preserves the gauge symmetry if the 5D fields transform in the following way:
\begin{eqnarray}
\hat{L}_M(x,z)&\to & e^{i\hat{\alpha}_{L}(x,z)}[\hat{L}_M(x,z)+i\partial_M]e^{-i\hat{\alpha}_{L}(x,z)},\nonumber\\
\hat{R}_M(x,z)&\to & e^{i\hat{\alpha}_{R}(x,z)}[\hat{R}_M(x,z)+i\partial_M]e^{-i\hat{\alpha}_{R}(x,z)},\nonumber\\
X(x,z)&\to & e^{i\hat{\alpha}_L(x,z)}X(x,z)e^{-i\hat{\alpha}_R(x,z)},\nonumber\\
N_{1,2}(x,z)&\to & e^{i\hat{\alpha}_{L,R}(x,z/\xi)}N_{1,2}(x,z),
\end{eqnarray}
where $\hat{L}_M=L^a_M t^a$, etc.

To be consistent, the metric for the nucleon sector should also be slightly modified: we should use
\begin{equation}\label{f_B}
f_B(z)^2=1-\frac{z^4}{(\xi z_h)^4}
\end{equation}
in the metric~(\ref{metric})~(that is, replace $z_h$ by $\xi z_h$), so that nucleons and mesons melt together at temperature $T=1/(\pi z_m)$.\footnote{The nucleons melt when the eigenvalue equation (\ref{Nucleon_eigen_eqn2}) becomes singular, that is, when the zero point of $f_B(z)$ lies between 0 and $z_B$: $0<\xi z_h\leq z_B$, or $T\geq 1/(\pi z_m)$. The use of a different $f(z)$~(and therefore a different geometry) in the nucleon sector seems counterintuitive, but this is avoided in the other attempt given in the appendix, where we use the same geometry for both sectors. Their finite-temperature predictions prove to be totally identical. }

The chiral symmetry breaking part of the Lagrangian is introduced as
\begin{equation}\label{Yukawa_12}
 S_\mathrm{Yukawa}=-g_{1/2}\int \mathrm{d}^4 x\int^{z_B}_{\epsilon}dz[\bar{N}_1 X(x,z/\xi) N_2+\mathrm{H.c.}].
\end{equation}
The spectrum of the nucleon sector is obtained~\cite{baryon_spectra,Nucleon_reso,baryon_T} by solving the eigenvalue equation that follows from action~(\ref{action_N}) and (\ref{Yukawa_12}):
\begin{eqnarray}\label{Nucleon_eigen_eqn2}
 \left(f_B\partial_z-\frac{\Delta_{+}}{z}\right)f_{1L} +\frac{g_{1/2} X_0(z/\xi)}{z}P~ f_{1R}=-\frac{M}{f_B(z)}f_{1R}, \nonumber \\
 \left(f_B\partial_z-\frac{\Delta_-}{z}\right)f_{1R}+\frac{g_{1/2} X_0(z/\xi)}{z}P~ f_{1L}=\frac{M}{f_B(z)}f_{1L},
\end{eqnarray}
where $\Delta_{\pm}=(f_B+\frac{1}{f_B})\pm m_5$, $M$ is the mass eigenvalue and $P$ is the parity of the state. The boundary condition is $f_{1R}(\epsilon)=f_{1R}(\xi z_m)=0$.

 There are two more parameters $\xi,~g_{1/2}$ in this sector; fitting to experimental data, we use $\xi=2.94,~g_{1/2}=71$. The calculated spectrum is listed in
Table~\ref{table:nucleon_spectrum}. Notice that since the calculated spectrum corresponds to the pole positions of the 2-point correlation function, we should compare our calculation to the pole masses listed in Ref.~\cite{PDG} rather than Breit-Wigner masses which are more commonly referred to.

\begin{table}[tbp]\centering
\caption{\label{table:nucleon_spectrum}Numerical results for nucleon masses at zero temperature}
\begin{tabular}{c|ccccc}
  \hline
  Nucleon Resonances & $p$ & $N(1440)$ & $N(1535)$ & $N(1650)$ & $N(1710)$  \\
  \hline
  $M_{\mathrm{exp}}$(GeV)~\cite{PDG} & $0.94$ &$1.37 $& $1.51$ & $1.66$& $1.72$\\
  $M_{\mathrm{th}}$(GeV)  & $0.97$ & $1.45$ & $1.29$ & $1.69$ & $1.84$  \\
  $J^P$& $\frac{1}{2}^+$&$\frac{1}{2}^+$&$\frac{1}{2}^-$&$\frac{1}{2}^-$&$\frac{1}{2}^+$\\
  Error & $3.2\%$ &$5.8\%$&$ 14.6\%$&$1.8\%$&$7.0\%$ \\
  \hline
\end{tabular}

\end{table}

\subsection{Spin-3/2 sector}
\label{subsec:spin32}
\subsubsection{The set-up}
Our convention mainly follows Ref.~\cite{S23}.
To study baryons with spin 3/2 and isospin 3/2, we introduce a pair of bulk Rarita-Schwinger fields $\Psi^M_{1abc}$
and $\Psi^M_{2\dot{a}\dot{b}\dot{c}}$ corresponding to 4D operators $O^{(\Delta)\mu}_{Labc}$ and $O^{(\Delta)\mu}_{R\dot{a}\dot{b}\dot{c}}$ respectively.  The SU(2)$_{L(R)}$ indices $a,b,c=1,2$~($\dot{a},\dot{b},\dot{c}=1,2$) are symmetric under permutations, forming the $J_L=3/2,J_R=0$~($J_L=0,J_R=3/2$) irreducible representation of SU(2)$_L\times$SU(2)$_R$: $\Delta_\mu^{++}=O^{(\Delta)}_{\mu 222},\Delta_\mu^{+}=O^{(\Delta)}_{\mu 221}$, etc.
Following the 4D Rarita-Schwinger Lagrangian ${\cal L}_{RS}=i\bar{\psi}_\mu\gamma^{\mu\rho\sigma}\partial_\rho\psi_\sigma-m\bar{\psi}_\mu\gamma^{\mu\sigma}\psi_\sigma$,
one can construct the 5D action for $\Psi^M_1$ and $\Psi^M_2$ as
\begin{eqnarray}\label{action_Psi}
S_{\Delta}=\int \mathrm{d}^4 x\int^{z_B}_{\epsilon}\mathrm{d}z &\sqrt{g}&\left[i\bar{\Psi}_{1A} \Gamma^{ABC}\nabla_B\Psi_{1C}-m_1\bar{\Psi}_{1A} \Psi_1^A-m_2\bar{\Psi}_{1A}\Gamma^{AB}\Psi_{1B}\right.\nonumber\\
&+& \left. (\Psi_1\rightarrow \Psi_2 \,\, ,\,\, L\rightarrow R  \,\, ,\,\, m_{1,2}\rightarrow-m_{1,2})\right],
\end{eqnarray}
where $\Psi_A=e_A^M \Psi_M$ and $\Gamma^A=(\gamma^m,-i\gamma^5)$ are the 5D gamma matrices, and $e_A^M$ denotes the vielbein satisfying $g_{MN}=e_M^A e_N^B \eta_{AB}$.
We also use notations $\Gamma^{ABC}=\Gamma^{[A}\Gamma^B\Gamma^{C]}/3!=(\Gamma^B\Gamma^C\Gamma^A-\Gamma^A\Gamma^C\Gamma^B)/2$ and $\Gamma^{AB}=[\Gamma^A,\Gamma^B]/2$. We use the same $z_B=\xi z_m$ as in the nucleon sector. The covariant derivative acting on $\Psi_{1,2}$ is defined similar to Eq.~(\ref{cov_D}), with $t^a$ replaced by the generator for $J=3/2$ representation
\begin{equation}\label{gen_rep32}
T^{a(3/2)}_{bb';cc';dd'}=t^a_{bb'}\delta_{cc'}\delta_{dd'}+t^a_{cc'}\delta_{dd'}\delta_{bb'}+t^a_{dd'}\delta_{bb'}\delta_{cc'}.
\end{equation}
For notational simplicity, we often omit the SU(2)$_L\times$SU(2)$_R$ indices, as we do in Eq.~(\ref{action_Psi}), keeping in mind that whenever we write something like $\bar{\Psi}^\mu T^a \Psi_\mu$, we mean $\bar{\Psi}^\mu_{bcd}T^{a(3/2)}_{bb';cc';dd'} \Psi_{\mu b'c'd'}$.

The chiral symmetry breaking part is a little more complex than that in the nucleon sector. Requiring invariance under gauge group SU(2)$_R\times$ SU(2)$_L$, the leading coupling term is given by
\begin{equation}\label{Yukawa_coupling}
{\cal L}_\mathrm{Yukawa}=-g_{3/2}\bar{\Psi}^{M }_{1\dot{a}\dot{b}\dot{c}}(x,z)X_{\dot{a}a}(x,z/\xi)X_{\dot{b}b}(x,z/\xi)X_{\dot{c}c}(x,z/\xi)\Psi_{2Mabc}(x,z)+\mathrm{H.c.}  .
\end{equation}

\subsubsection{The eigenvalue problem for spin-3/2 baryon spectrum}
For simplicity, we first derive the equations without Yukawa coupling. The $\Psi^M_1$ and $\Psi^M_2$ are then decoupled, so we simply write $\Psi^M$ for $\Psi^M_1$. For $\Psi^M_2$, just replace $m_{1,2}\to-m_{1,2}$. Extremizing the 5D action~(\ref{action_Psi}), we get the Rarita-Schwinger equation for $\Psi_A$:
\begin{equation}
i\Gamma^{ABC} \nabla_B\Psi_C-m_1\Psi^A-m_2\Gamma^{AB}\Psi_B=0,
\end{equation}
which can then be simplified to
\begin{equation}
i\Gamma^A \left(\nabla_A\Psi_B-\nabla_B\Psi_A\right)-m_-\Psi_B+\frac{m_+}{3}\Gamma_B \Gamma^A\Psi_A=0,
\end{equation}
where $m_{\pm}=m_1\pm m_2$.
The vielbein is defined as $g_{MN}=e_M^A e_N^B \eta_{AB}$, so we can choose it as
\begin{equation}\label{vielbein}
e_M^A=\frac{1}{z}\mathrm{diag}\left\{f_B(z),1,1,1,\frac{1}{f_B(z)}\right\}.
\end{equation}
The spin-connection can be worked out through its
definition $\omega_M^{AB}=-e^{KB}\partial_M e_K^A+e_L^A e^{KB}\Gamma_{KM}^L$:
\begin{eqnarray}\label{spin_connection}
\omega_0^{50}&=&-\omega_0^{05}=-\frac{2-f_B^2(z)}{z},\nonumber\\
\omega_i^{5i}&=&-\omega_i^{i5}=-\frac{f_B}{z}.
\end{eqnarray}
Other components we do not list here are zero. At zero temperature, $f_B(z)=1$, we see that Eq.~(\ref{spin_connection}) is consistent with that in previous works~\cite{baryon_spectra,nucleon_FF}. (The minus signs on the right-hand side of Eq.~(\ref{spin_connection}) are due to a different convention).

Being a reducible vector-spinor, the Rarita-Schwinger field contains not only a spin-$3/2$ component but spin-$1/2$ components as well. To project out the extra spin-$\frac{1}{2}$ components we introduce the following Lorentz-covariant constraint:
\begin{equation}\label{covariant_constraint}
e_A^M\Gamma^A\Psi_M=0.
\end{equation}

Putting the vielbein and spin-connection into the Rarita-Schwinger equation and using the above constraint, we get the following equations
\begin{eqnarray}\label{eom_Psi}
&iz\partial\!\!\!\!\!/\Psi_0-i\left(f_B(z)+\frac{1}{f_B(z)}\right)\Gamma^5\Psi_0-i\omega_0^{50}z f_B(z)\Gamma^0\Psi_5=m_-\Psi_0,\nonumber\\
&iz\partial\!\!\!\!\!/\Psi_i-i\left(f_B(z)+\frac{1}{f_B(z)}\right)\Gamma^5\Psi_i-i f_B(z)^2\Gamma^i\Psi_5=m_-\Psi_i,\nonumber\\
&iz\partial\!\!\!\!\!/\Psi_5-i\left(4f_B(z)-\frac{1}{f_B(z)}\right)\Gamma^5\Psi_5-ie^M_A\Gamma^A\partial_5\Psi_M=m_-\Psi_5,
\end{eqnarray}
(Note: the indices $0,i,5$ under $\Psi$ are local Lorentz coordinates) where
\begin{equation}\label{partial_T}
\partial\!\!\!\!\!/\equiv\frac{1}{z}e^M_A\Gamma^A\partial_M=\frac{1}{f_B(z)}\Gamma^0\partial_0+\Gamma^k\partial_k+f_B(z)\Gamma^5\partial_5.
\end{equation}
The difficulty in dealing with Eq.~(\ref{eom_Psi}) is that $\Psi_5$ and $\Psi_{0,i}$ get mixed. To settle this, we set $\Psi_5=0$~(we will justify this choice) which, combined with Eq.~(\ref{covariant_constraint}) and Eq.~(\ref{eom_Psi}), gives constraints
\begin{equation}\label{psi_constraints}
\Psi_0=0,~~~~~~\Gamma^i\Psi_i=0,
\end{equation}
and equation for $\Psi_i$:
\begin{equation}\label{eom_Psi_i}
iz\partial\!\!\!\!\!/\Psi_i-i\left(f_B(z)+\frac{1}{f_B(z)}\right)\Gamma^5\Psi_i=m_-\Psi_i.
\end{equation}

Now take chiral symmetry breaking into account, the Yukawa coupling term is given in Eq.~(\ref{Yukawa_coupling})~(replace $X_{\dot{a}a}=X_0(z)\delta_{\dot{a}a}$)
\begin{equation}\label{Yukawa_32}
{\cal L}_\mathrm{Yukawa}=-g_{3/2}\bar{\Psi}_{1M} X_0(z/\xi)^3 \Psi_2^M+\mathrm{H.c.},
\end{equation}
where $X_0(z)$ is given in Eq.~(\ref{exp:X}).
Then Eq.~(\ref{eom_Psi_i}) is modified to
\begin{eqnarray}\label{RS_EQ}
iz\partial\!\!\!\!\!/\Psi_{1i}-i\left(f_B(z)+\frac{1}{f_B(z)}\right)\Gamma^5\Psi_{1i}&=&m_-\Psi_{1i}+g_{3/2} X_0(z/\xi)^3\Psi_{2i},\nonumber\\
iz\partial\!\!\!\!\!/\Psi_{2i}-i\left(f_B(z)+\frac{1}{f_B(z)}\right)\Gamma^5\Psi_{2i}&=&-m_-\Psi_{2i}+g_{3/2} X_0(z/\xi)^3\Psi_{1i}.
\end{eqnarray}
As usual, we do the chirality decomposition
\begin{equation}\label{chiral_decomp}
\Psi^A_{r L,R}=\int \mathrm{d}^4p~F_{r L,R}(p,z)\psi^A_{L,R}(p) e^{-ip\cdot x},\quad\quad A=0, 1, 2, 3,\quad\quad r=1,2,
\end{equation}
where $\psi^A_{L,R}(p)=(1\pm\gamma^5)\psi^A(p)$ and $\psi^A(p)$  is defined as the solution of 4D Dirac equation:
\begin{equation}\label{4D_Dirac}
\left(\gamma^\mu p_\mu-M_{\Delta}\right)\psi^A(p)=0.
\end{equation}

It is important to notice that at finite temperature, the $x^0$ and $x^i$ coordinates in Eq.~(\ref{RS_EQ}) are not symmetric due to the $f_B(z)$ term in Eq.~(\ref{partial_T}). As is well known, in a finite temperature field theory, the time interval is finite (equals the inverse of temperature), thus breaking manifest Lorentz symmetry. As a result, the combination $(p^0)^2-|\vec{p}|^2$ does not appear as a whole after fourier transform, thus one should be careful in defining the boundary mass. In this paper we only focus on static particles, setting all spatial momentum to zero $p^i=0$ and the rest mass is recognized as the time component of 4-momentum $M=p^0$.

Thus, Eq.~(\ref{4D_Dirac}) with $p^\mu=(M_\Delta,0,0,0)$, combined with constraint (\ref{psi_constraints}) gives
\begin{equation}\label{Free_solution}
\psi^0(p)=0,\quad\quad \psi^i(p)=\left(
                                   \begin{array}{c}
                                     \xi^i \\
                                     \xi^i \\
                                   \end{array}
                                 \right),\quad\quad\mbox{where}~~\sigma^i\xi^i=0
\end{equation}
(we use chiral basis for Dirac matrices)~where $\sigma^i$ are the Pauli matrices. We immediately recognize $\xi^i$ as the vector-spinor representation of SO(3), with the $\sigma^i\xi^i=0$ constraint projecting out the extra spin-1/2 component in $1\otimes\frac{1}{2} =\frac{3}{2}\oplus\frac{1}{2}$. We can choose the $\vec{\xi}=\xi^i\vec{e}_i$ as
\begin{eqnarray}\label{xi_i}
&\vec{\xi}_{S_z=3/2}=\vec{z}_+ \zeta_\uparrow,\quad\quad\vec{\xi}_{S_z=1/2}=\sqrt{\frac{1}{3}}~\vec{z}_+ \zeta_\downarrow+\sqrt{\frac{2}{3}}~\vec{z}_0 \zeta_\uparrow,\nonumber\\
&\vec{\xi}_{S_z=-1/2}=\sqrt{\frac{2}{3}}~\vec{z}_0 \zeta_\downarrow+\sqrt{\frac{1}{3}}~\vec{z}_- \zeta_\uparrow,\quad\quad\vec{\xi}_{S_z=-3/2}=\vec{z}_- \zeta_\downarrow,
\end{eqnarray}
where
\begin{equation}
\vec{z}_0=\vec{z}\quad\quad\vec{z}_{\pm}=\mp\frac{\vec{x}\pm \mathrm{i}\vec{y}}{\sqrt{2}}\quad\quad\zeta_{\uparrow,\downarrow}=\left(
                                                                                                                   \begin{array}{c}
                                                                                                                     1 \\
                                                                                                                     0 \\
                                                                                                                   \end{array}
                                                                                                                 \right),\left(
                                                                                                                              \begin{array}{c}
                                                                                                                                0 \\
                                                                                                                                1 \\
                                                                                                                              \end{array}
                                                                                                                            \right).
\end{equation}
 Thus we see that the constraints (\ref{covariant_constraint}) and $\Psi_5=0$ indeed give us the four physical degrees of freedom for a static spin-3/2 particle.

Substituting Eq.~(\ref{chiral_decomp}) into Eq.~(\ref{RS_EQ}), with $p^i=0$, we get a set of eigenvalue equations
\begin{eqnarray}\label{Delta_eigen_eqn}
\left(
  \begin{array}{cc}
    f_B\partial_z-\frac{\Delta_{+}}{z} & -\frac{g_{3/2} X_0(z/\xi)^3}{z} \\
    -\frac{g_{3/2} X_0(z/\xi)^3}{z} & f_B\partial_z-\frac{\Delta_-}{z} \\
  \end{array}
\right)\left(
         \begin{array}{c}
           F_{1L} \\
           F_{2L} \\
         \end{array}
       \right)&=&-\frac{M_\Delta}{f_B(z)}\left(
                      \begin{array}{c}
                        F_{1R} \\
                        F_{2R} \\
                      \end{array}
                    \right),\nonumber\\
\left(
  \begin{array}{cc}
    f_B\partial_z-\frac{\Delta_-}{z}& \frac{g_{3/2} X_0(z/\xi)^3}{z} \\
    \frac{g_{3/2} X_0(z/\xi)^3}{z} & f_B\partial_z-\frac{\Delta_+}{z} \\
  \end{array}
\right)\left(
         \begin{array}{c}
           F_{1R} \\
           F_{2R} \\
         \end{array}
       \right)&=&\frac{M_\Delta}{f_B(z)}\left(
                      \begin{array}{c}
                        F_{1L} \\
                        F_{2L} \\
                      \end{array}
                    \right).
\end{eqnarray}
where $\Delta_{\pm}=(f_B+\frac{1}{f_B})\pm m_-$, $f_B$ is a shorthand for $f_B(z)$.

The boundary conditions are the same as those of the zero temperature case:
\begin{equation}
F_{1R}(\epsilon)=F_{1R}(\xi z_m)=F_{2L}(\xi z_m)=F_{2L}(\epsilon)=0.
\end{equation}
It can be proved that~\cite{Nucleon_reso,proof_parity} for parity-even states we have $F_{1L}=F_{2R}$ and $F_{2L}=-F_{1R}$ while for parity-odd states, we have $F_{1L}=-F_{2R}$ and $F_{2L}=F_{1R}$. Therefore, for a state with parity $P$, Eq.~(\ref{Delta_eigen_eqn}) can be simplified to
\begin{eqnarray}\label{Delta_eigen_eqn2}
 \left(f_B\partial_z-\frac{\Delta_{+}}{z}\right)F_{1L} +\frac{g_{3/2} X_0(z/\xi)^3}{z}P~ F_{1R}=-\frac{M_\Delta}{f_B(z)}F_{1R}, \nonumber \\
 \left(f_B\partial_z-\frac{\Delta_-}{z}\right)F_{1R}+\frac{g_{3/2} X_0(z/\xi)^3}{z}P~ F_{1L}=\frac{M_\Delta}{f_B(z)}F_{1L},
\end{eqnarray}
with boundary condition $F_{1R}(\epsilon)=F_{1R}(\xi z_m)=0$.

In addition to $z_m,~\sigma,~m_q,~\xi$, there are two more parameters $g_{3/2},~m_-$ in this equation. For $m_-$, the AdS/CFT prescription gives $|m_-|=|m_1-m_2|=\Delta_{3/2}-2$, where $\Delta_{3/2}$ is the conformal dimension of the $\Delta$ baryon interpolating field corresponding to $\Psi^\mu$. If we use its classical dimension $\Delta_{3/2}=9/2$ which leads to $|m_-|=5/2$, it turns out that the resulting spectrum does not match the experiment data. We argue that there is also an anomalous dimension contributing to $\Delta_{3/2}$ in strongly coupled QCD; however, since we do not know how to calculate it, we simply take it as an additional free parameter fitted to reproduce the $\Delta$ baryon spectrum.  The numerical fitting leads to $g_{3/2}=375,~m_-=8$, and the calculated spectrum is listed in Table~\ref{table:delta_spectrum}.
\begin{table}[tbp]\centering
\caption{$\label{table:delta_spectrum}$Numerical results for $\Delta$ masses at zero temperature }
\begin{tabular}{c|ccccc}
  \hline
  $\Delta$ Resonances & $\Delta(1232)$ & $\Delta(1600)$ & $\Delta(1700)$ & $\Delta(1920)$ & $\Delta(1940)$  \\
  \hline
  $M_{\mathrm{exp}}$(GeV)~\cite{PDG} & $1.21$ &$1.51$& $1.65$ & $1.90$& $1.94$\\
  $M_{\mathrm{th}}$(GeV)  & $1.17$ & $1.61$ & $1.62$ & 2.02 & 2.04  \\
  $J^P$& $\frac{3}{2}^+$&$\frac{3}{2}^+$&$\frac{3}{2}^-$&$\frac{3}{2}^+$&$\frac{3}{2}^-$\\
  Error & $3.3\%$ &$6.6\%$&$ 1.8\%$&$6.3\%$&$5.2\%$ \\
  \hline
\end{tabular}
\end{table}

\subsection{Pion-Baryon coupling}
\label{subsec:pbcoupling}
The calculation of the pion-nucleon coupling $g_{\pi N N}$ and pion-$\Delta$ coupling $g_{\pi\Delta\Delta}$ at zero temperature in the unitarity gauge can be found in Refs.~\cite{baryon_spectra,S23}. Since the meson sector is often studied in $A_5=0$ gauge, as we do in Sect.~\ref{subsec:boson}, here we present the calculation of these two quantities in $A_5=0$ gauge. Different choice of gauges proves to be completely equivalent. We mainly focus on $g_{\pi\Delta\Delta}$, the derivation of $g_{\pi NN}$ is similar.

\subsubsection{Expression of $g_{\pi\Delta\Delta}$ at finite temperature in $A_5=0$ gauge}
The $g_{\pi\Delta\Delta}$ effective coupling constant is obtained from the $\Delta$-pion coupling terms in the actions~(\ref{action_Psi}) and (\ref{Yukawa_coupling}):
\begin{eqnarray}\label{lpdd}
  {\cal L}^{(0)}_{\pi\Delta\Delta}&=&\frac{1}{z^4} [\bar{\Psi}^\mu_1 \Gamma^\nu \hat{A}_\nu(x,z/\xi) \Psi_{1\mu}-\bar{\Psi}^\mu_2 \Gamma^\nu \hat{A}_\nu(x,z/\xi) \Psi_{2\mu}]\nonumber\\
&\sim& \frac{1}{z^4} (\bar{\Psi}^\mu_1 \Gamma^\nu \partial_\nu\hat{\phi} \Psi_{1\mu}-\bar{\Psi}^\mu_2 \Gamma^\nu \partial_\nu\hat{\phi} \Psi_{2\mu})\nonumber\\
&=&-  \frac{1}{z^4}[\bar{\Psi}^i_1 (\overleftarrow{\partial\!\!\!\!\!/}\hat{\phi}+\hat{\phi}\overrightarrow{\partial\!\!\!\!\!/}) \Psi_{1i}-\bar{\Psi}^i_2 (\overleftarrow{\partial\!\!\!\!\!/}\hat{\phi}+\hat{\phi}\overrightarrow{\partial\!\!\!\!\!/}) \Psi_{2i}]\nonumber\\
&=&[-i\frac{f_B}{z^4}\bar{\Psi}^i_{1}~(\overleftarrow{\partial_z}\hat{\phi}+\hat{\phi}\overrightarrow{\partial_z})\gamma^5\Psi_{1i}+
2i\frac{1}{z^5}\bar{\Psi}^i_{1}\hat{\phi}\gamma^5(f_B+\frac{1}{f_B})\Psi_{1i}\nonumber\\
&&+\frac{i g_{3/2}X_0(z/\xi)^3}{z^5}(\bar{\Psi}^i_{1}\hat{\phi}\Psi_{2i}-\bar{\Psi}^i_{2}\hat{\phi}\Psi_{1i})]-(1\to 2)\nonumber\\
&=&-i\partial_z[\frac{f_B}{z^4}\bar{\Psi}^i_{1}~T^a\gamma^5\Psi_{1i}-(1\to 2)]\phi^a+\frac{2i g_{3/2}X_0(z/\xi)^3}{z^5}(\bar{\Psi}^i_{1}\hat{\phi}\Psi_{2i}-\bar{\Psi}^i_{2}\hat{\phi}\Psi_{1i})\nonumber\\
&=&i\frac{f_B}{z^4}[\bar{\Psi}^i_{1}~\partial_z\hat{\phi}~\gamma^5\Psi_{1i}-(1\to 2)]+\frac{2i g_{3/2}X_0(z/\xi)^3}{z^5}(\bar{\Psi}^i_{1}\hat{\phi}\Psi_{2i}-\bar{\Psi}^i_{2}\hat{\phi}\Psi_{1i}),
\end{eqnarray}
where $\hat{A}_\nu=A^a_\nu T^a$, the $\nu$ index in the first two lines is implicitly contracted by $\mathrm{diag}(1/f_B,-1,-1,-1)$, in the second line we use Eq.~(\ref{def_phi_pi}) and omit the irrelevant $A_{\nu\perp}$ term, in the third line we integrate by part, in the fourth line we use Eq. (\ref{RS_EQ}), and in the final line we integrate by part again.~\footnote{In the third line of Eq.~(\ref{lpdd}) we replace the $\mu$ index by a spatial index $i$, and do so similarly in Eq.~(\ref{lcmb}). This is not necessary at zero temperature, when the theory is totally covariant. At finite temperature, however, the theory is no longer covariant, and we only focus on static $\Delta$ particles whose $\Psi_0=\Psi_5=0$. }
\begin{eqnarray}\label{lykw}
 {\cal L}_\mathrm{Yukawa}&=&-\frac{1}{z^5}g_{3/2}\bar{\Psi}^{M }_{1\dot{a}\dot{b}\dot{c}}(x,z)X_{\dot{a}a}(x,z/\xi)X_{\dot{b}b}(x,z/\xi)X_{\dot{c}c}(x,z/\xi)\Psi_{2Mabc}(x,z)+\mathrm{H.c.}\nonumber\\
 =&&-\frac{1}{z^5}g_{3/2}\bar{\Psi}^{M }_{1\dot{a}\dot{b}\dot{c}}X_0(z/\xi)^3 2i\pi^k(x,z/\xi)(t^k_{\dot{a}a}\delta_{\dot{b}b}\delta_{\dot{c}c}+t^k_{\dot{b}b}\delta_{\dot{a}a}\delta_{\dot{c}c}+t^k_{\dot{c}c}\delta_{\dot{a}a}\delta_{\dot{b}b})\Psi_{2Mabc}+\mathrm{H.c.}\nonumber\\
=&&-\frac{2i g_{3/2}X_0(z/\xi)^3}{z^5}\bar{\Psi}^{M}_1\hat{\pi}(x,z/\xi)\Psi_{2M}+\mathrm{H.c.},
\end{eqnarray}
where we use $X_{\dot{a}a}(x,z)=[X_0(z)\exp(2i\hat{\pi})]_{\dot{a}a}=X_0(z)[\delta_{\dot{a}a}+2i\pi^k(x,z)t^k_{\dot{a}a}]$.

Combining Eqs.~(\ref{lpdd}) and (\ref{lykw}) we get
\begin{eqnarray}\label{lcmb}
{\cal L}_{\pi\Delta\Delta}=i\frac{f_B}{z^4}[\bar{\Psi}^i_{1}~\partial_z\hat{\phi}~\gamma^5\Psi_{1i}-(1\to 2)]+\frac{2i g_{3/2}X_0(z/\xi)^3}{z^5}(\bar{\Psi}^i_{1}T^a\Psi_{2i}-\bar{\Psi}^i_{2}T^a\Psi_{1i})(\phi^a-\pi^a).
\end{eqnarray}
The $g_{\pi \Delta\Delta}$ coupling constant is defined by the 4D effective Lagrangian~\footnote{Usually, vector-spinor representation is used for isospin $I=3/2$ representation and $g_{\pi\Delta\Delta}$ is defined as
$
{\cal L}_{\pi\Delta\Delta}=g_{\pi \Delta\Delta}\bar{\psi}^\mu_i i\gamma^5 \vec{\sigma}\cdot\vec{\pi}\psi_{\mu i}
$. The details of vector-spinor representation can be found in Ref.~\cite{Zhushilin}, for example, $\psi^\mu_+\equiv-(\psi^\mu_1+i\psi^\mu_2)/\sqrt{2}=(\Delta^{\mu++},\Delta^{\mu+}/\sqrt{3})^T
 $. With this, it is easy to check that this definition is equivalent to Eq.~(\ref{def_gpdd}).
  }
\begin{eqnarray}\label{def_gpdd}
L_{\pi\Delta\Delta}=ig_{\pi\Delta\Delta}\bar{\psi}^\mu_{\dot{a}bc}\gamma^5 (\vec{\pi}\cdot\vec{\sigma})_{\dot{a}a}\psi_{\mu abc},\nonumber\\
=i\frac{2}{3}g_{\pi\Delta\Delta}\bar{\psi}^\mu\gamma^5\vec{\pi}\cdot \vec{T}^{(3/2)}\psi_\mu.
\end{eqnarray}
Kaluza-Klein reducing the 5D fields $\Psi_{1,2},\pi,\phi$, using Eq.~(\ref{chiral_decomp}) we get
the expression for $g_{\pi\Delta\Delta}$:~\footnote{The power of $z$ in Eqs.~(\ref{exp:piDD}) and (\ref{norm_32}) is different from that in Ref.~\cite{S23}, due to the different definition of $\Delta$ wave functions $F_{L,R}$. We emphasize again that the index $A$ in Eq.~(\ref{chiral_decomp}) of our paper is a local Lorentz index. }
\begin{eqnarray}\label{exp:piDD}
g_{\pi\Delta\Delta}&=&\int_0^{z_B} \left\{\frac{-3\partial_z\phi(z/\xi) f_B}{2 z^4}\left(F^*_{1L}F_{1R}-F^*_{2L}F_{2R}\right)\right.\nonumber\\
&&\left.+\frac{3g_{3/2}v(z/\xi)^3[\phi(z/\xi)-\pi(z/\xi)] }{8z^5}\left(F^*_{1L}F_{2R}-F^*_{2L}F_{1R}\right)\right\} \mathrm{d}z.
\end{eqnarray}

\subsubsection{Expression for $g_{\pi NN}$  at finite temperature in $A_5=0$ gauge}

The derivation of $g_{\pi NN}$ effective coupling constant is similar. It is defined by the 4D effective Lagrangian:
 \begin{equation}\label{Lpnn_4D}
L_{\pi NN}=2i g_{\pi NN}\bar{\psi}\gamma^5\pi^a t^a\psi,
\end{equation}
 and is obtained from the nucleon-pion coupling terms in the actions~(\ref{action_N}) and (\ref{Yukawa_12})~\footnote{The factor $z f_B$ comes from the vielbein (\ref{vielbein}). Here $\Gamma^z=-i\gamma^5$ and the index ``z'' in $A_z$ is a general coordinate. }
\begin{equation}
{\cal L}_{\pi NN}=z f_B(\bar{N}_1 \Gamma^\mu A^a_\mu t^a N_1-\bar{N}_2 \Gamma^\mu A^a_\mu t^a N_2)-g_{1/2}(\bar{N}_1 X N_2+\mathrm{H.c.}).
\end{equation}
After Kaluza-Klein reducing the 5D fields~(parallel to Eqs.~(\ref{lpdd})--(\ref{exp:piDD})) , we get
\begin{eqnarray}\label{exp:piNN}
g^{(0)}_{\pi NN}&=&\int_0^{z_B} \left\{\frac{-\partial_z\phi(z/\xi)f_B}{2 z^4}\left(f^*_{1L}f_{1R}-f^*_{2L}f_{2R}\right)\right.\nonumber\\
&&\left.+\frac{g_{1/2} v(z/\xi)}{2 z^5} [\phi(z/\xi)-\pi(z/\xi)]\left(f^*_{1L}f_{2R}-f^*_{2L}f_{1R}\right)\right\}\mathrm{d}z.
\end{eqnarray}
For nucleons, we must add another term to the 5D action, as done in Ref.~\cite{Vega:2011prd,Vega:2012prd,nucleon_FF,Liu:2015jna},
\begin{equation}\label{L_FNN}
{\cal L}_{FNN}=i \eta_V [\bar{N}_1 \Gamma^{MN}F^{(L)}_{MN}(x,z/\xi)N_1-(1\to 2,L\to R)].
\end{equation}
This term is very important in explaining the nucleon anomalous magnetic moments; without this term, the calculated magnetic moment of proton would be $\mu_N$ rather than experimental value $2.79\mu_N$. The anomalous magnetic moments of proton and neutron derived from this term are~(in unit of $\mu_N=e/2m_N$)~\cite{Vega:2011prd,Vega:2012prd,nucleon_FF}
\begin{equation}\label{mu_ano}
\mu^{\rm{(ano)}}_{\rm{p,n}}=\mp 2\eta_V \int_{0}^{z_B}\frac{dz}{z^3} 2m_N f^*_{1L} f_{2L}.
\end{equation}
For proton, $\mu^{\rm{(ano)}}_{\rm{p}}=1.79$ and for neutron $\mu^{\rm{(ano)}}_{\rm{n}}=-1.91$, so we fit $|\mu^{\rm{(ano)}}_{\rm{p,n}}|=(1.79+1.91)/2=1.85$ which gives $\eta_V=0.330$.

The term~(\ref{L_FNN}) also contributes to pion-nucleon coupling $g_{\pi NN}$:
\begin{eqnarray}\label{g_pnn_1}
  g^{(1)}_{\pi NN} =&-&\eta_V\xi\int_0^{z_B}\frac{dz}{z^4}\partial_z\phi(z/\xi)\left[2 m_5 f_B(z)(f^*_{1L}f_{1R}-f^*_{2L}f_{2R})\right.\nonumber\\
&-& \left. f_B(z)^2 z(\partial_z f^{*}_{1L}f_{1R}+\partial_z f^{*}_{2L}f_{2R}-f^{*}_{1L}\partial_z f_{1R}-f^{*}_{2L}\partial_zf_{2R})\right.\nonumber\\
&+&\left. g_{1/2}f_B(z)v(z/\xi)(f^*_{1L}f_{2R}+f^*_{2L}f_{1R}) \right],
\end{eqnarray}
which should be added to expression~(\ref{exp:piNN}) to give the complete formula for $g_{\pi NN}$:
\begin{equation}
g_{\pi NN}=g^{(0)}_{\pi NN}+g^{(1)}_{\pi NN} .
\end{equation}

The normalizations of nucleon and $\Delta$ wave functions are fixed by requiring canonical kinetic terms in Eqs.~(\ref{action_N}) and (\ref{action_Psi}), respectively:~\footnote{At finite temperature it is not possible to write the relevant terms covariantly. For nucleons, ``canonical kinetic term'' means $i\bar{\psi} \gamma^0\partial_0 \psi$, this choice is the same as that in the zero temperature case for static particles. For spin-3/2 sector ``canonical kinetic term'' means $i\bar{\psi}_i \gamma^{i0j}\partial_0\psi_j$, as we have already pointed out in Eq.~(\ref{xi_i}) that $\psi_0=0$ for static spin-3/2 particles.}
\begin{equation}\label{norm_12}
\int_0^{z_B} \frac{\mathrm{d}z}{z^4 f_B(z)}\left(|f_{1L}|^2+|f_{2L}|^2\right)=1,
\end{equation}
\begin{equation}\label{norm_32}
\int_0^{z_B} \frac{\mathrm{d}z}{z^4 f_B(z)}\left(|F_{1L}|^2+|F_{2L}|^2\right)=1.
\end{equation}

\subsubsection{Numerical results for pion-baryon transition couplings}
With all these prescriptions at hand, we are able to numerically calculate the coupling constants. The results are shown in Table~\ref{table:couplings}.
\begin{table}[tbp]\centering
\caption{\label{table:couplings}Numerical results for pion-baryon effective couplings}
\begin{tabular}{c|cccccc}
  \hline
    & $g_{\pi NN}$& $g^{(1440)}_{\pi NN^*}$ &$g^{(1535)}_{\pi NN^*}$ & $g_{\pi\Delta\Delta}$ &$g^{(1600)}_{\pi\Delta\Delta^*}$ &$g^{(1700)}_{\pi\Delta\Delta^*}$\\
   \hline
  Exp & 13.2~\cite{data_gpnn} & 5.5--8.0($\approx$6.6) &0.56--0.83($\approx$0.70) &12--24~\footnotemark &10--19($\approx14$) &1.4--2.8($\approx2.1$)\\
  Th & 14.5 &5.7 & 0.95& 12&4.4 &2.0\\
  Error & 9.8\% &-&-& - &- &-\\
  \hline
\end{tabular}
\end{table}
\footnotetext{For $g_{\pi\Delta\Delta}$, the experimental data~\cite{g_pDD_data} give the range 1.1--30, and the predictions of other models~(such as $1/N_c$ expansion~\cite{g_pDD_largeNc1,g_pDD_largeNc2}, QCD sum rule~\cite{Zhushilin}, and QCD parameterization method~\cite{g_pDD_QCDparam}) varies from 12 to 24. }
The experimental values of transition coupling constants $g^{(1440)}_{\pi NN^*}$, $g^{(1535)}_{\pi NN^*}$, $g^{(1600)}_{\pi\Delta\Delta}$, $g^{(1700)}_{\pi\Delta\Delta}$ are obtained from experimental decay rates. The decay rate formula of $N(1440)$ and $N(1535)$ can be found in Ref.~\cite{Decaywidth,Decaywidth2}:
\begin{equation}\label{Nucleon_decay_formula}
\Gamma(N^*\to N\pi)=3\frac{g^2_{\pi NN^*}}{4\pi m_{N^*}}|\vec{p}|(E-P m_N)
\end{equation}
where $P$ is the parity of the excited state $N^*$, $E$ is the energy of the final nucleon and $|\vec{p}|$ is the center-of-mass momentum. The connection between $f_{\pi NN^*}$ and $g_{\pi NN^*}$ is $g_{\pi NN^*}=\frac{m_{N^*}+ P m_N}{m_\pi}f_{\pi NN^*}$.

The partial decay rate of a $\Delta^*$ excited state with parity $P$ is
\begin{equation}\label{Delta_decay_formula}
\Gamma(\Delta^* \to \Delta(1232)\pi)=\frac{5g_{\pi\Delta\Delta^*}^2}{36\pi m_{\Delta^*}}|\vec{p}|(\frac{2 E^3}{3m^2_\Delta}+\frac{E}{3}-P m_\Delta),
\end{equation}
where $E$ is the energy of $\Delta(1232)$ baryon.

We see that except $g^{(1600)}_{\pi\Delta\Delta^*}$, all of them are consistent with experiment. Maybe $g^{(1600)}_{\pi\Delta\Delta^*}$ could get better if we introduce a term similar to (\ref{L_FNN}), however, as there is no data on the magnetic moment of $\Delta$ baryons to fix the $\eta_V$ parameter, we do not consider it in this paper.

\subsection{Summary of results at zero temperature }
\label{subsec:sumzero}
We summarize the free parameters and the calculated observables at zero temperature in Tables~\ref{table:parameters} and \ref{table:summary}. The 5D mass of nucleons $m_5=5/2$ is fixed by AdS/CFT prescription, so it is not listed.
\begin{table}[tbp]\centering
 \caption{\label{table:parameters}Free parameters in our model.}
\begin{tabular}{cccccccc}
  \hline
  $z_m^{-1}$ & $m_q$ & $\sigma^{1/3}$ & $\xi$&$g_{1/2}$ &  $g_{3/2}$ & $m_-$ & $\eta_V$\\
  \hline
  346~MeV & 2.30~MeV & 308~MeV &2.94& 71 & 375 & 8 & 0.330\\
  \hline
  \end{tabular}
  \end{table}
  \begin{table}
  \caption{\label{table:summary}Summary of results at zero temperature.}
  \begin{tabular}{c|ccccccc}
  \hline
  Meson & $m_\pi$ & $f_\pi$ & $m_\rho$ &$F^{1/2}_\rho$ & $m_{a_1}$ & $F^{1/2}_{a_1}$&\\
  \hline
  Exp~(MeV) & 139.6 & 92.4 & 775.3 & 329 & 1230 &433&\\
  Th~(MeV)  & 141   & 84.0 & 832   & 353 & 1220 &440&\\
  Error& 1.0\%&9.1\% & 7.3\% &7.3\%&0.8\%&1.6\%&\\
  \hline
  Nucleons & $p$ & $N(1440)$ & $N(1535)$ & $N(1650)$ & $N(1710)$ & \multicolumn{1}{|c}{$\mu_p$}  &$\mu_n$ \\
  \hline
  $M_{\mathrm{exp}}$(GeV) & $0.94$ &$1.37 $& $1.51$ & $1.66$& $1.72$ &\multicolumn{1}{|c}{ 2.79~$\mu_N$}  &-1.91~$\mu_N$\\
  $M_{\mathrm{th}}$(GeV)  & $0.97$ & $1.45$ & $1.29$ & $1.69$ & $1.84$&\multicolumn{1}{|c}{2.85~$\mu_N$} &-1.85~$\mu_N$ \\
  Error & $3.2\%$ &$5.8\%$&$ 14.6\%$&$1.8\%$&$7.0\%$ &\multicolumn{1}{|c} {2.2\%} & 3.1\%\\
  \hline
  $\Delta$ Resonances & $\Delta(1232)$ & $\Delta(1600)$ & $\Delta(1700)$ & $\Delta(1920)$ & $\Delta(1940)$ & & \\
  \hline
  $M_{\mathrm{exp}}$(GeV) & $1.21$ &$1.51$& $1.65$ & $1.90$& $1.94$& & \\
  $M_{\mathrm{th}}$(GeV)  & $1.17$ & $1.61$ & $1.62$ & 2.02 & 2.04 & & \\
  Error & $3.3\%$ &$6.6\%$&$ 1.8\%$&$6.3\%$&$5.2\%$ & &\\
  \hline
  Couplings & $g_{\rho\pi\pi}$ & $g_{\pi NN}$& $g^{(1440)}_{\pi NN^*}$ &$g^{(1535)}_{\pi NN^*}$ & $g_{\pi\Delta\Delta}$ &$g^{(1600)}_{\pi\Delta\Delta^*}$ &$g^{(1700)}_{\pi\Delta\Delta^*}$\\
   \hline
  Exp &6.03 & 13.2 & 5.5--8.0 &0.56--0.83 & 12--24 &10--19 &1.4--2.8\\
  Th &5.29 & 14.5 &5.7 & 0.95& 12 &4.4 &2.0\\
  Error &12.3\% & 9.8\% &-&-& - &- &-\\
  \hline
\end{tabular}
\end{table}
We see that with these parameters, the calculated spectrum and coupling constants fit well with experiment~(except that the error of $N(1535)$ is somewhat bigger), the rms error of all the observables~(with clear experimental data) being remarkably 6.7\%.

\section{Predictions at Finite Temperature}
\label{sec:finite}
\begin{figure}[tbp]\centering
\includegraphics[width=\textwidth]{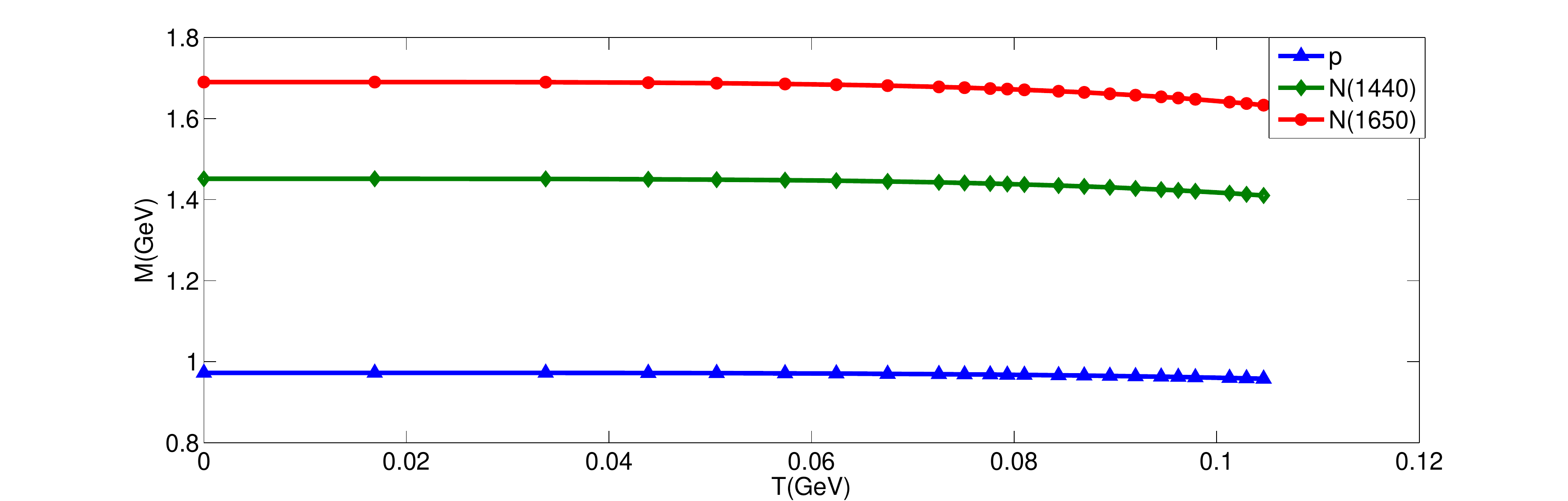}
\caption{\label{fig:N_T}The temperature dependence of nucleon masses.}
\end{figure}

\begin{figure}[tbp]\centering
\includegraphics[width=\textwidth]{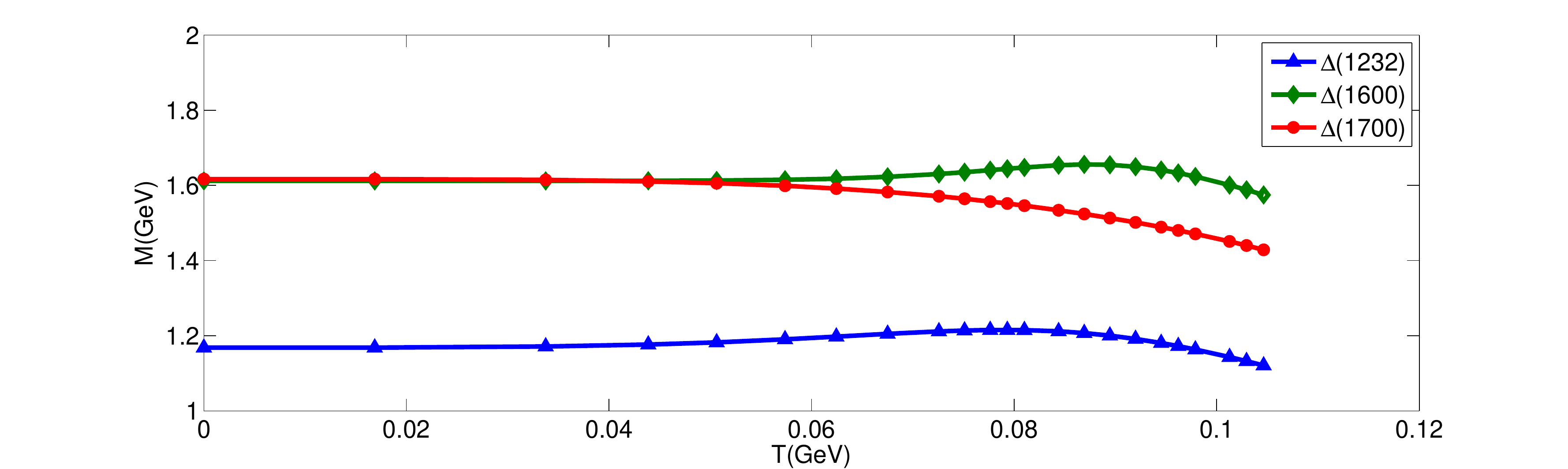}
\caption{\label{fig:Delta_T}The temperature dependence of $\Delta$ masses.}
\end{figure}

Once the parameters $z_m$, $\sigma$, $m_q$, $\xi$, $m_-$, $g_{1/2}$, $g_{3/2}$, $\eta_V$ are fixed, we are about to study the temperature dependence of the observables listed in Table~\ref{table:summary}. To do this, use the same eigenvalue equations and formulae for coupling constants, setting $f(z)=1-(z/z_h)^4$ and $f_B(z)=1-(z/\xi z_h)^4$. In Figure~\ref{fig:Delta_T} we draw the temperature dependence of the masses of the $\Delta(1232)$, $\Delta(1600)$, $\Delta(1700)$ baryons. By comparison, we also draw the temperature dependence of the nucleon sector in Figure~\ref{fig:N_T}, which is reminiscent of a previous work~\cite{baryon_T}.\\

We see that the spectrum of $\Delta$ baryons at finite temperature show some new features compared with nucleons: while the $\Delta(1700)$ resembles the nucleon sector, the masses of $\Delta(1232)$ and $\Delta(1600)$ increase a little bit before dropping down near the transition temperature. The behavior of $\Delta(1232)$ mass at low temperature qualitatively agrees with the result obtained by effective field theory calculations~\cite{Delta_T} and preliminary data from STAR collaboration~\cite{STAR_2004}.

We expect these results to be useful in studying thermal pion-nucleon scattering. In heavy ion collision experiments~(such as RHIC experiments), the collision of heavy nuclei at ultrarelativistic energies produces such a high energy density that for a short
time the system is in local thermal equilibrium and can be regarded as a finite temperature environment. The nucleons and virtual pions in the original nuclei collide each other, creating $\Delta$ baryons as intermediate resonance states. Thus a knowledge of the $\Delta$ baryon mass at nonzero temperature will help us understand these experiments better. For example, we see that at temperature around 0.1~GeV, the mass of $\Delta(1700)$ drop well below the mass of $\Delta(1600)$, so that the decay process $\Delta(1600)\to \Delta(1700)+\pi$, which is kinematically forbidden at zero temperature, can happen at that temperature.

\begin{figure}[tbp]\centering
\includegraphics[width=\textwidth]{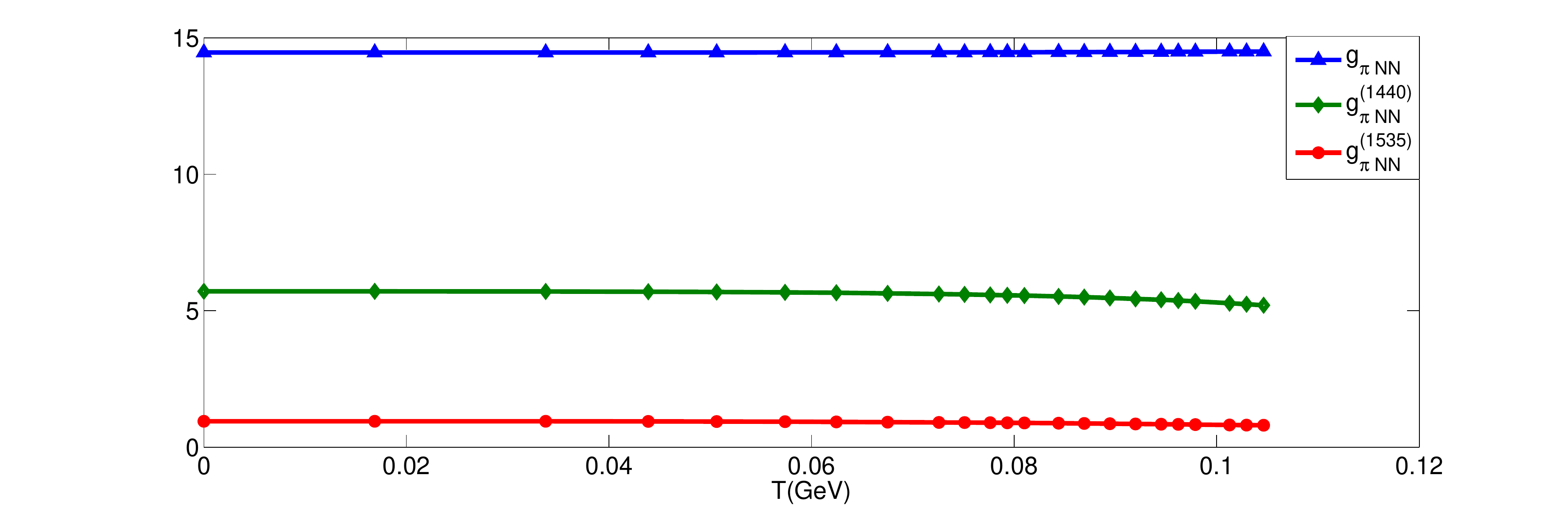}
\caption{\label{fig:gpnnT}The temperature dependence of pion-nucleon couplings.}
\end{figure}

\begin{figure}[tbp]\centering
\includegraphics[width=\textwidth]{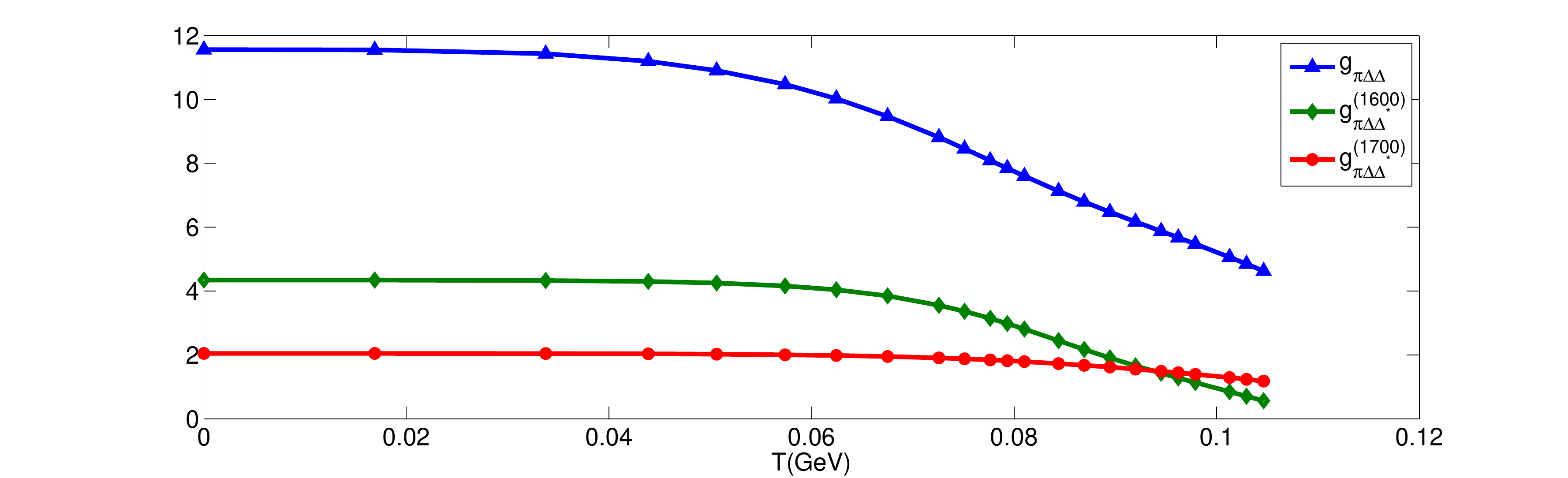}
\caption{\label{fig:gpddT}The temperature dependence of pion-$\Delta$ couplings.}
\end{figure}

The temperature dependence of the couplings ($g_{\pi NN}$ and $g_{\pi\Delta\Delta}$) can be obtained using Eqs.~(\ref{exp:piDD}), (\ref{exp:piNN}) and (\ref{g_pnn_1}). The results are shown in Figures~\ref{fig:gpnnT} and \ref{fig:gpddT}.

\begin{figure}[tbp]\centering
\includegraphics[width=\textwidth]{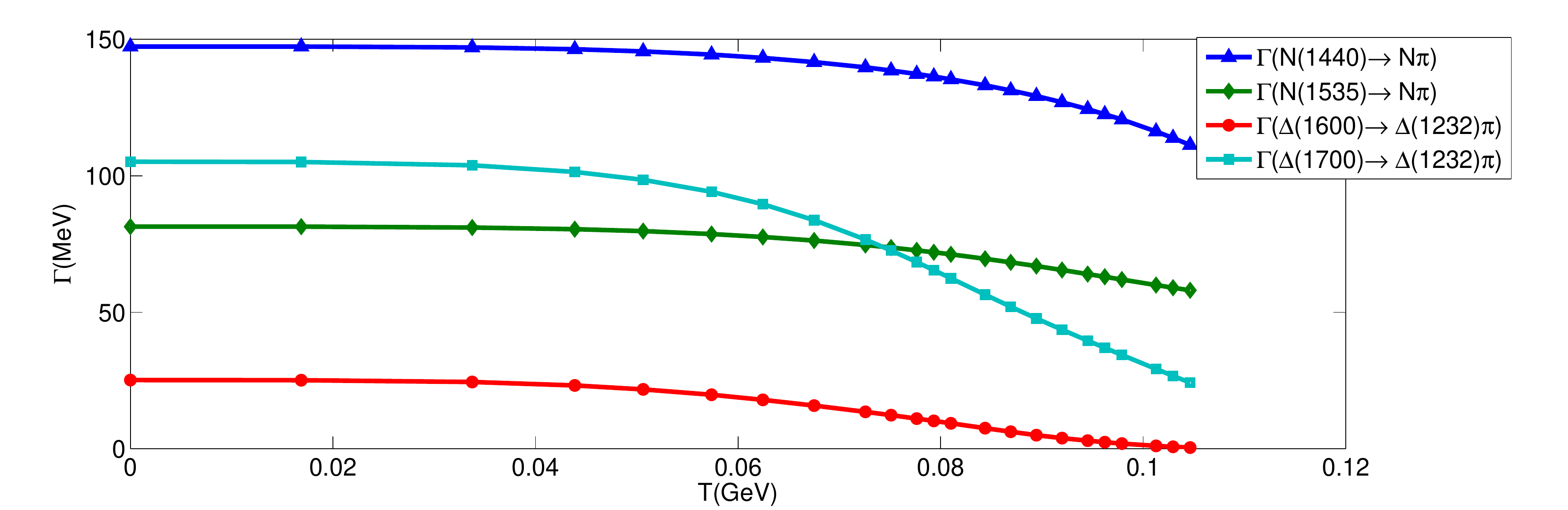}
\caption{\label{fig:GammaT}The temperature dependence of partial decay rates of excited states.}
\end{figure}

The temperature dependence of the transition couplings $g^{(1440)}_{\pi NN^*}$, $g^{(1535)}_{\pi NN^*}$, $g^{(1600)}_{\pi\Delta\Delta^*}$ and $g^{(1700)}_{\pi\Delta\Delta^*}$ can be directly checked by experiments. As they are related to the partial decay rate of the corresponding resonance states, we can make a prediction for the partial decay width of resonance states at finite temperature using Eqs.~(\ref{Nucleon_decay_formula}) and (\ref{Delta_decay_formula}). The result is shown in Figure~\ref{fig:GammaT}.
We see that they all tend to decrease as temperature increases, indicating that excited states tend to be more stable in a finite temperature nuclear environment. We hope that future experimental data at RHIC will check these predictions.

\section{Summary}
\label{sec:summary}
In this paper we make a unified approach to hadron properties at zero and nonzero temperature in AdS/QCD. By using a different IR cut-off and redefine the covariant derivative on baryon field in a self-consistent way, we combine the meson and baryon sectors and arrive at a universal holographic model of hadrons. The eigenvalue equations for mass spectrum and formulae for transition coupling constants are derived. Solving the model numerically, we fix the parameters to reproduce experimental data. The result is promising--using this set of parameters, all calculated observables with clear experimental data fit reasonably well, with rms error being remarkably 6.7\%.

We then turn the temperature on and get an interesting temperature dependence of the spectrum and couplings. Although our parameters are different from those used by previous works, the temperature dependence of nucleon spectrum reproduce earlier results. The $\Delta$ spectrum at finite temperature, however, shows some new features: while the $\Delta(1700)$ resembles the nucleon sector, the masses of $\Delta(1232)$ and $\Delta(1600)$ increase a little bit before dropping down near the transition temperature. The transition coupling constants $g_{\pi\Delta\Delta}$, $g^{(1600)}_{\pi\Delta\Delta^*}$ and $g^{(1700)}_{\pi\Delta\Delta^*}$ all decrease as temperature increases. We finally use these results to calculate the partial decay width of excited states at finite temperature, and give the prediction that these excited states tend to be more stable as temperature increases. We expect these results to be useful for studying thermal pion-nucleon scattering in finite temerature nuclear environment.

\begin{acknowledgements}
This work is supported by the National Natural Science Foundation of China (Grants No.~11035003, No.~11120101004 and No.~11475006) and by the National Fund for Fostering
Talents of Basic Science (Grant Nos.~J1103205 and J1103206). It is also supported by the Principal Fund for Undergraduate Research at Peking
University.
\end{acknowledgements}

\appendix
\section{Another way to introduce the scale magnification factor $\xi$}
We discuss in the introduction section about the difficulty that a single $z_m$ can not fit both sectors, and in the subsequent sections we resolve this difficulty by using different $z_m$ for the two sectors and somehow redefine the covariant derivative in a consistent way. We now present another way to resolve this difficulty. This one is relatively more technical, but it avoids the counterintuitive choice of using different geometries for the two sectors in Eq.~(\ref{f_B}). Their results for all calculated observables are the same. 

In this attempt we use the \textit{same} AdS-Sch background (\ref{metric}) with the \textit{same} IR cut-off $z_B=z_m=1/(346~\rm{MeV})$ for the baryon sector, for example, the 5D action for nucleons is
\begin{eqnarray}\label{action_Nnew}
S_{N}=\int \mathrm{d}^4 x\int^{z_m}_{\epsilon}dz &\sqrt{g}&\left[\frac{i}{2}\bar{N}_1 e^M_A\Gamma^A\nabla_M N_1- \frac{i}{2}(\nabla^\dagger_M\bar{N}_1) e^M_A\Gamma^A N_1-m_5\bar{N}_1 N_1\right.\nonumber\\
&&\left.+(1\rightarrow 2 \,\, ,\,\, L\rightarrow R  \,\, ,\,\, m_5\rightarrow-m_5)\right],
\end{eqnarray}
and similarly for $S_{\Delta}$~(also let $f_B(z)=f(z)$). Now, the value of $z_B$ is only $1/\xi$ times the one we previously used, so if we still follow the usual AdS/QCD method, we will get baryon masses $\xi$ times as big as before (notice $z_B$ has inverse mass dimension).  To avoid this, we modify the AdS/CFT prescription~(the correspondence between 5D fields and 4D operators) slightly: while $L^a_\mu(x,z)$ and $R^a_\mu(x,z)$ still correspond to $J^{a\mu}_{L,R}(x)$, the 5D nucleon field~($\Delta$-field) $N(x,z)$~($\Psi^\mu(x,z)$) corresponds to 4D operators $O(\xi x)$~($O^{(\Delta)}_\mu(\xi x)$) rather than $O(x)$~($O^{(\Delta)}_\mu(x)$), the boundary couplings are:
\begin{equation}\label{BC_new}
\int d^4x [L^{a(0)}_{\mu}(x)J^{a\mu}_{L}(x)+\bar{N}^{(0)}_1(x)O_L(\xi x)+
\bar{\Psi}^{\mu(0)}_{1}(x) O^{(\Delta)}_{L\mu}(\xi x)+(1\rightarrow 2 \,\, ,\,\, L\rightarrow R)],
\end{equation}
where the ``(0)'' indicates boundary value of 5D fields. As a result, the usual step of functionally differentiating on-shell action with respect to nucleon boundary fields yields $\langle0|TO(\xi x)\bar{O}(0)|0\rangle$ rather than $\langle0|TO(x)\bar{O}(0)|0\rangle\equiv G(x,0)$:
\begin{equation}\label{Gxix}
G_0(x,0)\equiv-i\frac{\delta^2 S_{5D}}{\delta \bar{N}^{(0)}(x)\delta N^{(0)}(0)}=\langle0|TO(\xi x)\bar{O}(0)|0\rangle= G(\xi x,0),
\end{equation}
and therefore,
\begin{equation}\label{Gpole}
\tilde{G}(p)\equiv\int d^4 x~ e^{ip\cdot x}G(x,0)=\int d^4 x~ e^{ip\cdot x}G_0(\frac{x}{\xi},0)=\xi^4\tilde{G}_0(\xi p),
\end{equation}
where $\tilde{G}_0(p)$ is the Fourier transform of $G_0(x,0)$--the 2-point function calculated from the usual step of functional differentiation. Thus, the pole spectrum of $\tilde{G}(p)$ is $1/\xi$ the pole spectrum of $\tilde{G}_0(p)$. This is exactly what we want, since the pole spectrum of $\tilde{G}_0(p)$ is $\xi$ times we previously obtained as a consequence of using $z_B=z_m$.

Thus, using the same IR cut-off in Eq.~(\ref{action_Nnew}) and in the mean time modifying the boundary coupling as done in (\ref{BC_new}) indeed give us the correct spectrum.

A potential danger of using the new boundary coupling (\ref{BC_new}) is that it may not give us the correct delta function of momentum conservation when we calculate the 3-point function $\langle0|T J^{a}_\mu(x) O(y) \bar{O}(z) |0\rangle$~(we may get something like $\delta^4(\xi(p'-p)-q)$). To solve this, we should define the 3-point coupling terms in the 5D action slightly differently. Specifically, we should modify the covariant derivative to
\begin{eqnarray}\label{cov_D_new}
\nabla_\mu&=&\partial_\mu+\frac{i}{4}\omega^{AB}_\mu\Sigma_{AB}-i\xi L^a_{\mu}(\xi x,z) t^a,\nonumber\\
\nabla_5&=&\partial_5+\frac{i}{4}\omega^{AB}_5\Sigma_{AB}-iL^a_5(\xi x,z) t^a.
\end{eqnarray}
and the Yukawa coupling to
\begin{equation}\label{Yukawa_new}
 S_\mathrm{Yukawa}=-g_{1/2}\int \mathrm{d}^4 x\int^{z_B}_{\epsilon}dz[\bar{N}_1(x,z) X(\xi x,z) N_2(x,z)+\mathrm{H.c.}].
\end{equation}
The gauge symmetry is still preserved if the 5D fields transform in the following way:
\begin{eqnarray}
\hat{L}_M(x,z)&\to & e^{i\hat{\alpha}_{L}(x,z)}[\hat{L}_M(x,z)+i\partial_M]e^{-i\hat{\alpha}_{L}(x,z)},\nonumber\\
\hat{R}_M(x,z)&\to & e^{i\hat{\alpha}_{R}(x,z)}[\hat{R}_M(x,z)+i\partial_M]e^{-i\hat{\alpha}_{R}(x,z)},\nonumber\\
X(x,z)&\to & e^{i\hat{\alpha}_L(x,z)}X(x,z)e^{-i\hat{\alpha}_R(x,z)},\nonumber\\
N_{1,2}(x,z)&\to & e^{i\hat{\alpha}_{L,R}(\xi x,z)}N_{1,2}(x,z),
\end{eqnarray}
As an example, we calculate the $g_{\pi NN}$ coupling constant at zero temperature in this new scheme to show that everything works all right.
The nucleon-pion coupling terms in the actions~(\ref{action_Nnew}) and (\ref{Yukawa_new}) are
\begin{eqnarray}
{\cal L}_{\pi NN}&=&z \xi[\bar{N}_1(x,z) \Gamma^\mu A^a_\mu(\xi x,z) t^a N_1(x,z)-(1\rightarrow 2)]-\frac{g_{1/2}}{z^5}[\bar{N}_1(x,z) X(\xi x,z) N_2(x,z)+\mathrm{H.c.}]\nonumber\\
&=&-\frac{1}{z^4}\phi(\xi x, z)\partial_\mu(\bar{N}_1\gamma^\mu N_1-\bar{N}_2\gamma^\mu N_2)-\frac{2ig_{1/2}}{z^5}\pi(\xi x,z)[\bar{N}_1 X_0(z) N_2-\bar{N}_2 X_0(z) N_1]\nonumber\\
&=&\frac{i\partial_z\phi(\xi x,z)}{z^4}(\bar{N}_1\gamma^5 N_1-\bar{N}_2\gamma^5 N_2)+\frac{2ig_{1/2}[\phi(\xi x,z)-\pi(\xi x,z)]}{z^5}[\bar{N}_1 X_0(z) N_2-\bar{N}_2 X_0(z) N_1],\nonumber\\
\end{eqnarray}
Kaluza-Klein reducing the 5D fields as usual, we get
\begin{eqnarray}
S_{\pi NN}&=&\int_0^{z_m} \xi^4\left\{\frac{-\partial_z\phi(z)}{2 z^4}\left(f^*_{1L}f_{1R}-f^*_{2L}f_{2R}\right)
+\frac{g_{1/2} v(z)}{2 z^5} [\phi(z)-\pi(z)]\left(f^*_{1L}f_{2R}-f^*_{2L}f_{1R}\right)\right\}dz\nonumber\\
&&\cdot\int d^4x \bar{N}^{(0)}(x)\gamma^5\pi^a(\xi x)\sigma^a N^{(0)}(x)/\xi^4.
\end{eqnarray}
The last factor is to be recognized as the 4D effective action. Since the 4D operator $O(x)$ corresponds to $N^{(0)}(x/\xi)/\xi^4$~(see Eq.~(\ref{BC_new})), we rewrite it as
\begin{equation}
\int d^4x \bar{N}^{(0)}(x)\gamma^5\pi^a(\xi x)\sigma^a N^{(0)}(x)/\xi^4=\int d^4x \frac{\bar{N}^{(0)}(x/\xi)}{\xi^4}\gamma^5\pi^a(x)\sigma^a \frac{N^{(0)}(x/\xi)}{\xi^4}.
\end{equation}
Thus, comparing with Eq.~(\ref{Lpnn_4D}), we get
\begin{equation}\label{gpnnnew}
g^{(0)}_{\pi NN}=\int_0^{z_m} \xi^4\left\{\frac{-\partial_z\phi(z)}{2 z^4}\left(f^*_{1L}f_{1R}-f^*_{2L}f_{2R}\right)
+\frac{g_{1/2} v(z)}{2 z^5} [\phi(z)-\pi(z)]\left(f^*_{1L}f_{2R}-f^*_{2L}f_{1R}\right)\right\}dz.
\end{equation}
The normalization of nucleon 5D fields should be fixed by requiring canonical kinetic terms in on-shell 5D action:
\begin{eqnarray}\label{norm_new}
S_{\mathrm{on-shell}}&=&\int d^5x \frac{1}{z^4}\left(|f_{1L}|^2+|f_{2L}|^2\right)\bar{N}^{(0)}(x)i\partial\!\!\!\!\!/N^{(0)}(x)\nonumber\\
&=&\int_0^{z_m} dz \frac{1}{z^4}\left(|f_{1L}|^2+|f_{2L}|^2\right)\xi^5 ~\int d^4x\frac{\bar{N}^{(0)}(x/\xi)}{\xi^4}i\partial\!\!\!\!\!/\frac{N^{(0)}(x/\xi)}{\xi^4},
\end{eqnarray}
so we should require
\begin{equation}\label{norm_new1}
\xi^5\int_0^{z_m} \frac{dz}{z^4}\left(|f_{1L}|^2+|f_{2L}|^2\right)=1.
\end{equation}
Comparing Eqs.~(\ref{gpnnnew}),~(\ref{norm_new1}) and Eqs.~(\ref{exp:piNN}),~(\ref{norm_12}), we find that they are equivalent if we identify
\begin{equation}\label{identify}
f_{1,2;L,R}(z)=\xi^4 \tilde{f}_{1,2;L,R}(\xi z),
\end{equation}
where $\tilde{f}_{1,2;L,R}$ is what we use in the main sections.

It is not difficult to check that every observables calculated in this scheme are equal to what we previously obtained, both at zero and finite temperatures.


\begin{thebibliography}{99}
\bibitem{Maldacena}
J.M.~Maldacena,
Adv.\ Theor.\ Math.\ Phys.\ {\bf 2} 231 (1998).
  arXiv:hep-th/9711200


\bibitem{AdS_CFT_review}
O.~Aharony et al.,
  Phys.\ Rept. {\bf 323} 183 (2000).
 arXiv:hep-th/9905111

\bibitem{QCD_conformal}
S.J.~Brodsky, G.F.~de Teramond,
\prd{77}{2008}{056007}.
arXiv:0707.3859 [hep-ph]

\bibitem{Brodsky:2014yha}
S.J.~Brodsky, G.F.~de Teramond, H.G.~Dosch, and J.~Erlich,
  Phys.\ Rep.\  {\bf 584}, 1 (2015).

\bibitem{meson_spectra_Brodsky}
S.J.~Brodsky, G.F.~de Teramond,
\plb{582}{2004}{211}.
arXiv:hep-th/0310227

\bibitem{Gutsche:2013prd}
T.~Gutsche, V.E.~Lyubovitskij, I.~Schmidt, A.~Vega,
  \prd{87}{2013}{056001}.
  arXiv:1212.5196 [hep-ph]

\bibitem{meson_spectra_Erlich}
J.~Erlich et al.,
\prl{95}{2005}{261602}.
arXiv:hep-ph/0501128


\bibitem{meson_EMFF}
S.J.~Brodsky, G.F.~de Teramond,
\prd{78}{2008}{025032}.
arXiv:0804.0452 [hep-ph]

\bibitem{meson_TFF}
F.~Zuo, Y.~Jia, T.~Huang,
\epjc{67}{2010}{253}.
arXiv:0910.3990 [hep-ph]

\bibitem{PiPiFF}
F.~Zuo, T.~Huang,
\epjc{72}{2011}{1813}.
arXiv:1105.6008 [hep-ph]

\bibitem{witten}
E.~Witten,
Adv.\ Theor.\ Math.\ Phys. {\bf 2} 505 (1998).
arXiv:hep-th/9803131

\bibitem{meson_T}
K.~Ghoroku, M.~Yahiro,
\prd{73}{2006}{125010}.
arXiv:hep-ph/0512289

\bibitem{vmeson_T}
M.~Fujita, K.~Fukushima, T.~Misumi, M.~Murata,
\prd{80}{2009}{035001}.
arXiv:0903.2316 [hep-ph]

\bibitem{Heavy1}
H.~Boschi-Filho, N.R.F.~Braga, C.N.~Ferreira,
\prd{74}{2006}{086001}.
arXiv:hep-th/0607038

\bibitem{Heavy2}
Y.~Kim, J.P.~Lee and S.H.~Lee,
\prd{75}{2007}{114008}.
arXiv:hep-ph/0703172

\bibitem{QGP_trans1}
C.P.~Herzog,
\prl{98}{2007}{091601}.
arXiv:hep-th/0608151

\bibitem{QGP_trans2}
K.~Kajantie, T.~Tahkokallio, J.T.~Yee,
\jhep{01}{2007}{019}.
arXiv:hep-ph/0609254

\bibitem{baryon_spectra}
D.K.~Hong, T.~Inami, H.~U.~Yee,
\plb{646}{2007}{165}.
arXiv:hep-ph/0609270


\bibitem{Zhangpeng}
P.~Zhang,
\prd{81}{2010}{114029}.
arXiv:1002.4352 [hep-ph]


\bibitem{unifyapproach}
H-C.~Kim, Y.~Kim, U.~Yakhshiev,
\jhep{11}{2009}{034}.
arXiv:0908.3406 [hep-ph]

\bibitem{S23}
H.C.~Ahn, D.K.~Hong, C.~Park, S.~Siwach,
\prd{80}{2009}{054001}.
arXiv:0904.3731 [hep-ph]

\bibitem{RHIC_1997}
E.L.~Hjort et al.,
\prl{79}{1997}{4345}.

\bibitem{STAR_2004}
P.~Fachini, 
 J.\ Phys.\ G {\bf 30} S735 (2004).
arXiv:nucl-ex/0403026.

\bibitem{v_meson_f}
L.A.H.~Mamani, A.S.~Miranda, H.~Boschi-Filho, N.R.F.~Braga,
\jhep{03}{2014}{058}.
arXiv:1312.3815 [hep-th]

\bibitem{spectra_HP_trans}
P.~Colangelo, F.~Giannuzzi, S.~Nicotri,
\prd{80}{2009}{094019}.
arXiv:0909.1534 [hep-ph]

\bibitem{Nucleon_reso}
T.~Gutsche, V.E.~Lyubovitskij, I.~Schmidt, A.~Vega,
  \prd{87}{2013}{016017}.
  arXiv:1212.6252 [hep-ph]

\bibitem{baryon_T}
Z.~Li, B.-Q.~Ma,
\prd{89}{2014}{015014}.
arXiv:1312.3451 [hep-ph]

\bibitem{PDG}
K.A.~Olive  et al.~(Particle Data Group),
Chin.\ Phys.\ C {\bf 38}, 090001 (2014).

\bibitem{nucleon_FF}
Z.~Abidin, C.E.~Carlson,
\prd{79}{2009}{115003}.
arXiv:0903.4818 [hep-ph]


\bibitem{proof_parity}
T.~Gutsche, V.E.~Lyubovitskij, I.~Schmidt, A.~Vega,
\prd{86}{2012}{036007}.
arXiv:1204.6612 [hep-ph]




\bibitem{Vega:2011prd}
A.~Vega, I.~Schmidt, T.~Gutsche, V.E.~Lyubovitskij,
  \prd{83}{2011}{036001}.
 arXiv:1010.2815 [hep-ph]

\bibitem{Vega:2012prd}
A.~Vega, I.~Schmidt, T.~Gutsche, V.E.~Lyubovitskij,
  \prd{85}{2012}{096004}.
 arXiv:1202.4806 [hep-ph]

\bibitem{Liu:2015jna}
  T.~Liu, B.-Q.~Ma,
  \prd{92}{2015}{096003}.
  arXiv:1510.07783 [hep-ph]

\bibitem{Decaywidth}
D.O.~Riska, G.E.~Brown,
\npa{679}{2001}{577}.
arXiv:nucl-th/0005049

\bibitem{Decaywidth2}
M.T.~Pena, D.O.~Riska, A.~Stadler,
\prc{60}{1999}{045201}.
arXiv:nucl-th/9902066

\bibitem{data_gpnn}
D.V.~Bugg,
\epjc{33}{2004}{505}.

\bibitem{Delta_T}
H.~van Hees, R.~Rapp,
\plb{606}{2005}{59}.
arXiv:nucl-th/0407050


\bibitem{Zhushilin}
S.-L.~Zhu,
\prc{63}{2000}{018201}.
arXiv:nucl-th/0009062







\bibitem{g_pDD_data}
R.A.~Arndt et al.,
\prd{20}{1979}{651}.


\bibitem{g_pDD_largeNc1}
R.F.~Dashen, E.~Jenkins, A.V.~Manohar,
\prd{49}{1994}{4713}.
arXiv:hep-ph/9310379

\bibitem{g_pDD_largeNc2}
R.F.~Dashen, E.~Jenkins, A.V.~Manohar,
\prd{51}{1995}{3697}.
arXiv:hep-ph/9411234

\bibitem{g_pDD_QCDparam}
A.J.~Buchmann, S.A.~Moszkowski,
\prc{87}{2013}{028203}.
arXiv:1304.2194 [hep-ph]
\end{thebibliography}
\end{document}